\documentclass{emulateapj}
\shorttitle{Properties of Far-Infrared Galaxies}
\shortauthors{M.Symeonidis et al.}
\begin{document}
\title{AEGIS: Infrared Spectral Energy Distributions of MIPS 70 $\mu m$ 
selected sources}
\author{M.~Symeonidis,$\!$\altaffilmark{1}
D.~Rigopoulou,$\!$\altaffilmark{1}
J.-S.\ Huang,$\!$\altaffilmark{2}
M. Davis,$\!$\altaffilmark{3}
M. L. N. Ashby,$\!$\altaffilmark{2}
P. Barmby,$\!$\altaffilmark{2}
E. Egami,$\!$\altaffilmark{4}
G. G. Fazio,$\!$\altaffilmark{2}
E. LeFloc'h, $\!$\altaffilmark{4}
G. Rieke, $\!$\altaffilmark{4}
S. P. Willner,$\!$\altaffilmark{2} 
G. Wilson,$\!$\altaffilmark{5}
}
\altaffiltext{1}{Astrophysics Department, Oxford University, Oxford OX1 
3RH}
\altaffiltext{2}{Harvard-Smithsonian Ctr. for Astrophysics, 60 Garden St.,
Cambridge, MA 02138}
\altaffiltext{3}{Dept. of Astron., Univ. of California Berkeley, Campbell Hall,
Berkeley, CA 94720}
\altaffiltext{4}{Steward Observatory, University of Arizona, 933 N. Cherry 
Avenue Tucson, AZ 85721}
\altaffiltext{5}{Spitzer Science Ctr., California Inst. of Technology, 
1200E California Blvd., Pasadena, CA 91125}

\begin{abstract}

We present 0.5 -- 160 $\mu$m Spectral Energy Distributions (SEDs) 
of galaxies, detected at 70$\mu$m with the Multiband Imaging Photometer 
for Spitzer (MIPS), using broadband imaging data from Spitzer 
and ground-based telescopes. 
Spectroscopic redshifts, in the range 0.2$\le$z$\le$1.5, have been measured
as part of the Deep Extragalactic Evolutionary Probe2 (DEEP2) project.
Based on the SEDs we explore the nature and physical properties 
of the sources. Using the optical spectra we derive H$_{\beta}$ and [OII]-based
Star Formation Rates (SFR) which are 10--100 times lower than SFR estimates
based on IR and radio. The median offset in SFR between optical and IR 
is reduced by 
a factor of $\sim$3 when we apply a typical extinction corrections. 
We investigate mid-to-far infrared correlations for low redshift ($>$0.5) and
high redshift (0.5$<$z$<$1.2) bins.
Using this unique ``far--infrared'' selected sample
we derive an empirical mid to far-infrared relationship that can be used to 
estimate 
the infrared energy budget of galaxies in the high-redshift universe. 
Our sample can be used as a template to translate
far-infrared luminosities into bolometric luminosities for high
redshift objects.

\end{abstract}


\section{Introduction}

The first breakthrough in infrared astronomy followed the launch of
the Infrared Astronomical Satellite (IRAS), which uncovered a
substantial population of galaxies very luminous in the infrared
(e.g. Neugebauer et al. 1984, Soifer et al. 1987). Its successor, the
Infrared Space Observatory (ISO), demonstrated that a significant
fraction of the energy budget in star-forming galaxies emerges in the
infrared regime (e.g Soifer et al. 1984). Since then, attempts to
describe the nature of these objects have lead to extensive studies of
the properties of dust and its presence in star forming regions, using
Spectral Energy Distributions (SEDs) and infrared colors as parameter
constraints (e.g. Rowan-Robinson $\&$ Crawford 1989, Dale $\&$ Helou
2002). At the same time, correlations using photometry at different
wavelengths have proved extremely useful in facilitating the study of
the high-z universe and compensating for lack of long wavelength
InfraRed (IR) data (e.g. Sanders $\&$ Mirabel 1996).

Since the launch of the Spitzer Space Telescope (SST) (Werner et
al. 2004), we are able to study the emission from interstellar dust
across a wide range of environments both locally and at high
redshifts. Deep surveys with the InfraRed Array Camera (IRAC) and the
Multiband Imaging Photometer for Spitzer (MIPS) (Rieke et al. 2004)
have provided a wealth of energy distributions at every redshift. This
letter presents the analysis of a sample of high-redshift sources
detected at 70$\mu$m by MIPS, in the region of the Extended Groth
Strip (EGS). We have used a multitude of photometry in the near, mid
and far infrared to construct SEDs and examine galaxy colours and have
demonstrated the compatibility of this population with samples of
local infrared luminous galaxies.

\section{The Observations}

The All-Wavelength Extended Groth Strip International Survey (AEGIS) is a
multi-Wavelength survey with a coverage of 2.0\degr$\times$10\arcmin (Davis
et al. 2006). MIPS observations of the Extended Groth Strip were conducted in 
two epochs: January and June of 2004. The effective EGS area with both IRAC and MIPS coverage is 725 arcsec$^{2}$. 180 70$\micron$ sources
were detected in this region with a point source sensitivity of 4 mJy 
at 5$\sigma$, hereafter the MIPS 70$\mu$m sample.

The IRAC and MIPS Basic Calibrated Data (BCD) delivered by the Spitzer Science 
Center (SSC) included flat-field and linearity corrections, dark subtraction 
and flux calibration. The BCD data were further processed by each team's own 
refinement routines, including distortion corrections, pointing refinement, 
mosaicing and cosmic ray removal by sigma-clipping. We used DAOPHOT to extract 
sources from both IRAC and MIPS images; a Point Spread Function (PSF) FWHM of 
1.8$\arcsec$--2.0$\arcsec$ for IRAC and 35$\arcsec$/40$\arcsec$ for 
MIPS 70/160 $\mu$m was used. The aperture fluxes in each band were 
subsequently corrected to total fluxes using known PSF growth curves 
from Fazio et al. (2004), Huang et al. (2004).

All 70$\micron$ sources in the sample are detected at the IRAC 3.6, 4.5,
5.8, 8.0 $\mu$m, and MIPS 24 $\micron$ bands. However, because of the different
alignment of the MIPS and IRAC images we do not have IRAC fluxes for
all 70$\mu$m sources.
88$\%$ of sources in the
sample are also detected at 160$\micron$. In addition , the AEGIS data
set (Davis et al. 2006) provides multi-wavelength photometry and
spectroscopic redshifts for the 70$\micron$ sample, which permit the
study of their X-ray-to-radio SEDs. 42 sources in the 70$\mu$m
population have a reliable DEEP2 spectroscopic redshift in the range
of $0<z<1.2$ (we only use redshifts with qualities 3 or 4, according to the 
DEEP2 classification). We further divide the 70$\mu$m sources in two redshift 
bins: $z \le 0.5$ and $0.5<z<1.2$ (the terms ``low'' and ``high'' redshift 
hereafter). Although, the DEEP2 project uses colour criteria to preselect 
objects with photometric redshifts in the range 0.7-1.55, these constraints
have not been imposed on the EGS, therefore our redshift sample is not
biased against the 70$\mu$m sample. We limit the analysis in this
letter to 38 objects for which we have full visible-IR SEDs and
reliable (see above) spectroscopic redshifts (the remaining 4 objects do not
have full SED coverage).

\section{Properties of the 70$\mu$m population}

\subsection{Spectral Energy Distributions}

Using CFHT B,R,I imaging data and Spitzer photometry (3.6, 4.5, 5.8,
8.0, 24, 70 and 160 $\mu$m) we were able to construct 96 full SEDs, with 
38 having reliable spectroscopic redshifts. The SEDs of all objects 
with reliable spectroscopic redshifts (38 sources) are shown in Figure 1.
We group the SEDs in three panels: objects with low-z (left), 
high-z (middle) and, a separate panel for objects with AGN-like 
SEDs (right).  

Objects from the low$/$high-z groups show typical galaxy SEDs: the
optical bands are dominated by starlight which is then thermally
reprocessed by dust and re-emitted at longer wavelengths, giving
rise to the noticeable increase in Mid-InfraRed (MIR) flux. It is not
surprising that the optical/near-IR part of the high-z SEDs is
almost an order of magnitude lower than the low-z equivalent, as it
samples a shorter wavelength region more strongly affected by
extinction. 

In the mid-infrared regime (ie. 12$<\lambda<$24 $\mu$m) we find that the
high-z objects have overall much higher flux densities than the low-z objects. 
There are two explanations for this: either high-z sources have higher hot
dust emission shortwards of 20$\mu$m
or, low-z sources have higher cold dust emission longwards of 60$\mu$m.
Low-z objects show a predominant excess flux in the IRAC 8$\mu$m band, which we
attribute to emission at 6.2 and/or 7.7$\mu$m, from Polycyclic
Aromatic Hydrocarbons (PAHs). 

In the far-infrared we note a large scatter in 160$\mu$m fluxes evident in
both low and high-z sources. 
We find 70$\mu$m/160$\mu$m $\nu f_{\nu}$ ratios of
$\sim$1 and $\sim$0.4 for the low and high-z sources,
respectively. This is in agreement with the energy distributions of
local HII region-like galaxies (e.g. Dale et al. 2005, hereafter
D05). In the Spitzer Infrared Nearby Galaxy Survey (SINGS) results,
presented in D05, approximately 35$\%$ of the objects show a turnover
at 160$\mu$m and their average  70/160 $\nu f_{\nu}$ ratios are
$\sim$1; accordingly, we evaluate our objects to be equally
``warm'' to local sources.

Overall, the majority of our objects have characteristics consistent
with those of luminous star-forming infrared galaxies. However, a
small fraction (9 out of 96) reveal power-law type SEDs indicative of
the presence of an Active Galactic Nucleus (AGN) (of the 9 AGN-candidates 
only 3 have secure DEEP2 spectroscopic redshifts). These
power law-type sources have distinctly redder IRAC colors and are
clearly identified in IRAC-color plots. Interestingly,
the available optical spectra do not show any of the characteristic
signatures of an AGN, such as broad wings in the Balmer$/$Oxygen
lines; it is only with the addition of MIR data that we are able to
peer through the obscuring dust and reveal the active nucleus. Only
1/3 of these candidate-AGNs are also detected in the  x-rays, not
surprising, as the Chandra full band (0.5-10 keV)  data (flux limit
3.5$\times 10^{-15}$ ergs/s/$cm^2$)  (Georgakakis et al. 2006) are not
as deep as the MIPS and IRAC detections.  The selected fraction of
power law-type galaxies agrees  with a similar study in Frayer et
al. (2006). However, taking into  account ultra-hard (5-10 keV) x-ray
fluxes and hardness ratios  leads us to an upper limit of 15$\%$ of
AGNs in the sample,  consistent with the findings of Fadda et
al. (2002),  Franceschini et al. (2003). Brandt et al. (2006), show
that in their  24$\mu$m survey the contribution of AGNs is high
($>40\%$) when  selecting the brightest z$>$0.6 sources, but
$\sim$10$\%$ when  considering the whole redshift range. Our results
are consistent  with this study, as our power law-type sources with
spectroscopic  redshifts are found at z$>$0.6.

\subsection{Star Formation Rates}
 
In this section we carry out a comparative study of different diagnostics in
order to trace the Star Formation Rate (SFR) in our sample. Such
a study is possible given the wealth of information available for the AEGIS
project (Davis et al. 2006). We derive SFRs based on optical Balmer/[OII] 
emission lines, IR (present work) and 20 cm radio (Ivison et al. 2006)
luminosities. Our computations are based on the prescriptions of 
Kennicutt (1998) and  Bressan, Silva $\&$ Granato (2002).

All available spectra were examined for the 
presence of hydrogen
recombination and oxygen forbidden emission lines, as tracers of young
stellar populations. We identified H$\beta$ and [OII] features in
22$\%$ and 10$\%$ of the spectra, respectively. We extracted line
fluxes by calibrating the continuum using CFHT optical/near-IR data
and included stellar absorption corrections for each Balmer line. 
We derive an ``average'' SFR$_{opt}$ of 2 M$_{\odot}/yr$.
Likewise we derived average SFR$_{IR}$ and SFR$_{radio}$ of 100 
and 80 M$_{\odot}/yr$, respectively.

In Figure 2 we show a comparison of optical vs IR vs radio SFRs as a 
function of redshift for 38 objects 
from the 70 $\mu$m sample, which have full IRAC and MIPS photometry 
as well as reliable
spectroscopic redshifts in the range $0<z<1.2$. 
Radio and IR SFRs reach
values of $\sim$900 $M_{\odot}/yr$, revealing an active star-forming
population. The SFR$_{IR}$ is on average 50 times higher than the
non-extinction-corrected SFR$_{opt}$. Assuming an A$_{V}\sim$2 (a typical A$_{V}$ value for infrared
selected objects, see e.g. Rigopoulou et al. 2000) then 
SFR$_{IR}/$SFR$_{optical} \sim $ 10 confirming that extinction is indeed very
high. The fact that SFR$_{IR} \sim$ SFR$_{radio}$ reinforces the notion that
the 70 $\mu$m population is indeed made up of dusty star-forming galaxies.
Moreover, it provides a nice confirmation that the radio--far-IR correlation
that was found to exist for local galaxies (e.g. Condon, Anderson \& Helou, 
1991) extends out to 
z$\sim$1 sources. Finally, it is important to stress that any attempts to 
derive SFR estimates based on optical$/$UV measurements are likely to provide 
severe underestimates.

\section{The mid-to-far infrared correlation}

In this section we investigate the mid-to-far infrared luminosity correlations.
We limit our study to objects with confirmed spectroscopic redshifts and full 
SEDs, and explore the correlations for each redshift bin separately. For our
computations we have adopted a cosmology with $\Omega_M$=0.3,
${\Omega}_{\Lambda}$=0.7 and $H_0=71 km sec^{-1}Mpc^{-1}$. 
Investigation of spectral energy distributions and
behavior of colors (see earlier text) has lead us to the conclusion
that our 70$\mu$m sample shares many similar characteristics with local
infrared luminous galaxies. Consequently to estimate K-corrections
we used the complete SED of the local galaxy NGC 4631 
(convolved to the Spitzer resolution) and followed the 
prescription of Hogg (2002). We are, of course, aware that use of a single
template to derive K-corrections may introduce some biases. However, as 
Appleton et al. (2004) also points out the presently available models do not 
fully encompass the observed parameters especially in the far-infrared. This is likely to change with the availability of more Spitzer far-infrared datasets
such as SINGS. 

In Figure 3 we plot the 24 vs 70 and 70 vs 160 $\mu$m correlations.
We have performed a weighted least-square fit for each case, taking
into account the calibration errors of 10 and 20$\%$ for the 24 and
70$\mu$m fluxes respectively, as well as the errors arising from SED
fitting. We quantified the strength of the correlations using the
Pearson linear coefficient. The Pearson
coefficient ranges between 0 and $|1|$  for no or excellent
correlation respectively, with any intermediate value indicating the
degree of linear dependence between the variables. Our values of
$\sim$0.8 for the 24--70$\mu$m plot and $\sim$0.9 for the
70--160$\mu$m plot indicate a strong relation. 

As expected, most of the energy in this population is emitted in the
far-IR, also supported by the lines of best fit. By combining the
above results we derived a relationship to estimate the monochromatic
luminosity at 160$\mu$m from data at 24 and 70$\mu$m, with an accuracy
higher than 30$\%$, for the low-z and high-z cases.
\begin{equation}
\label{eqn:corr1}
\textrm{For z} \le 0.5:
\qquad L_{160} \sim 0.9L_{24}+0.8L_{70}\qquad (L_{\odot} )
\end{equation}
\begin{equation}
\label{eqn:corr2}
\textrm{For z} > 0.5:
\qquad L_{160} \sim 0.6L_{24}+0.4L_{70}\qquad (L_{\odot})
\end{equation}

It is evident that the well-studied mid-to-far infrared relation found
to hold for local galaxies (e.g. Dale $\&$ Helou 2002) can also be
extended to the non-local universe. Although, by limiting our study to
the objects with spectroscopic redshifts, we select in favor of the
brightest sources, we still manage to sample a wide range of
luminosities representative of z$\sim$1 populations emerging through
various deep infrared surveys. The correlations we have found  
can be used to obtain a direct estimate of the far-infrared luminosity
of z$\sim$1 objects and avoid the large errors introduced from MID-FIR
extrapolations of various galaxy templates.

\acknowledgments{}

MS has been supported by a Marie Curie Excellence Grant while working on this
project MEXT-CT-2003-002792.
DR acknowledges support from the Leverhulme Trust via a Research Fellowship.
This work is based on observations made with the Spitzer Space
Telescope, which is operated by the Jet Propulsion Laboratory,
California Institute of Technology under NASA contract 1407. Support
for this work was provided by NASA through Contract Number 1256790
issued by JPL.

\begin{figure}
\begin{minipage}[t]{6cm}
\centering
\includegraphics[width=6cm]{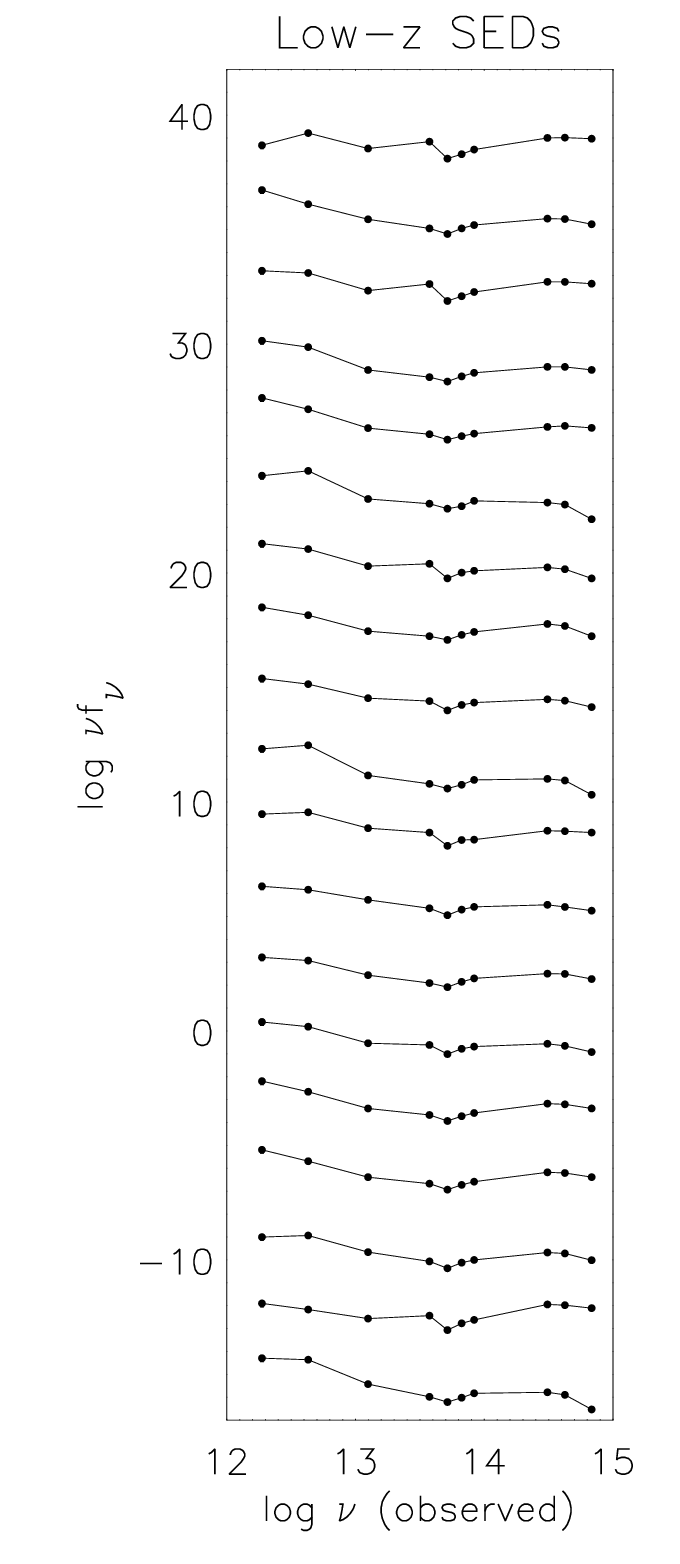}
\end{minipage}
\begin{minipage}[t]{6cm}
\centering
\includegraphics[width=6cm]{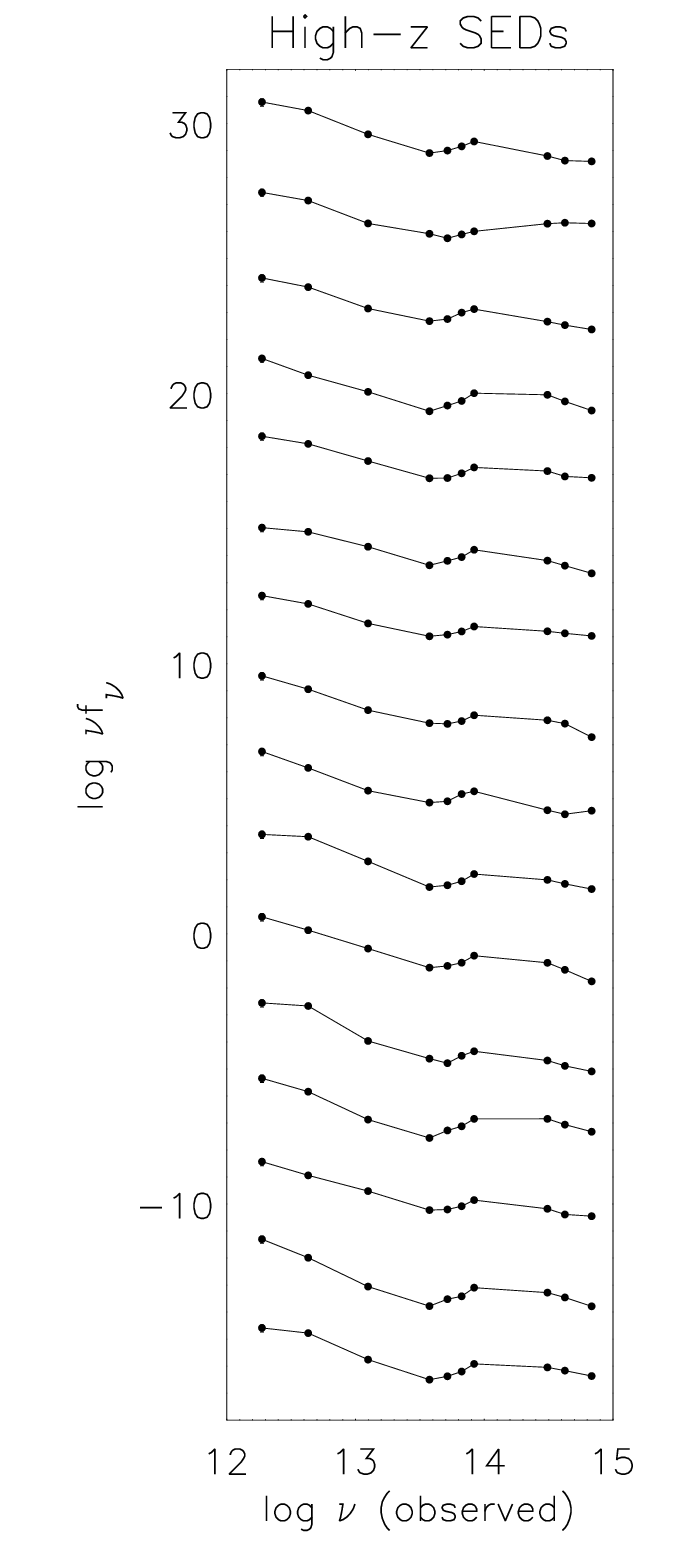}
\end{minipage}
\begin{minipage}[t]{6cm}
\centering
\includegraphics[width=6cm]{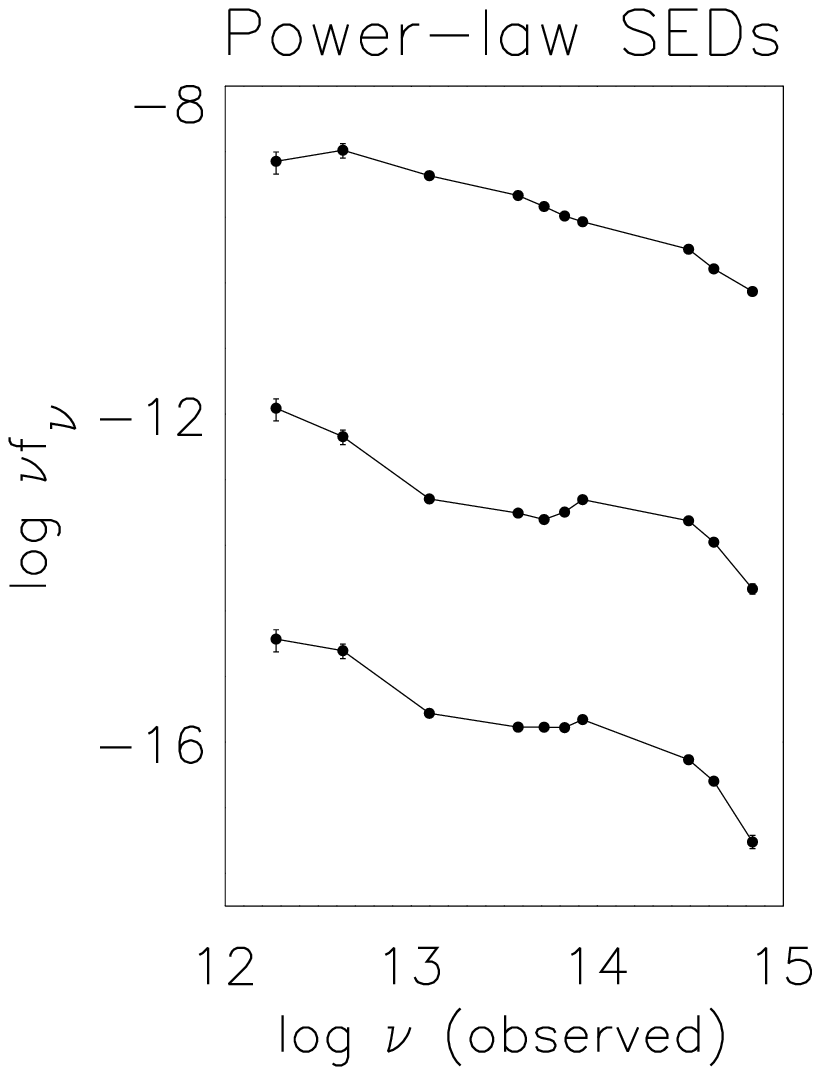}
\end{minipage}
\caption{Full SEDs for the entire MIPS 70$\mu$m sample with confirmed
spectroscopic redshifts grouped
into low-z (left), high-z (middle) and power-law (right) panels. 
The lower SED in each panel is in
real y-axis units - for clarity, each subsequent SED in the low-z and power-law groups is transposed by 3 dex; in the high-z group the SEDs are transposed by 2 dex. In the direction of decreasing frequency, the
10 point photometry used to assemble the SEDs is: B, R, I  magnitudes
from CFHT, 3.6, 4.5, 5.8, 8 $\mu$m fluxes from IRAC and 24, 70, 160
$\mu$m fluxes from MIPS}
\end{figure}
\clearpage

\begin{figure}
\centering
\includegraphics[width=10cm]{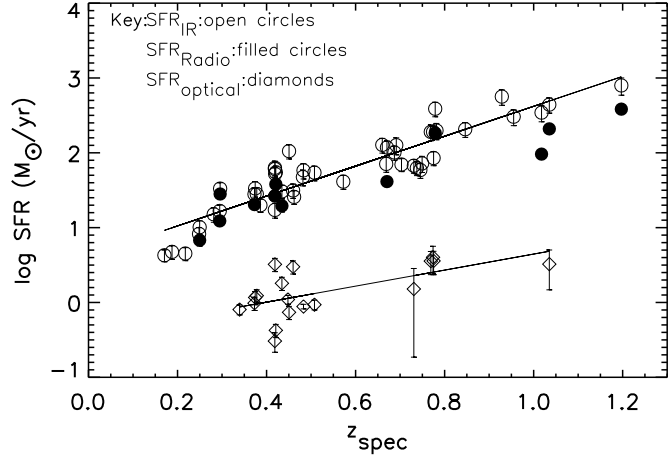}
\caption {Plot of various indicators of the Star Formation Rate (SFR) as 
a function of redshift. SFRs estimated based on optical (H$_{\beta}$ and [OII])
emission lines (open diamonds), radio (filled circles) and IR (open circles) 
luminosities. The line-fits represent the mean SFR$_{IR}$ and SFR$_{opt}$.}
\end{figure}

\begin{figure}
\centering
\includegraphics[width=14cm]{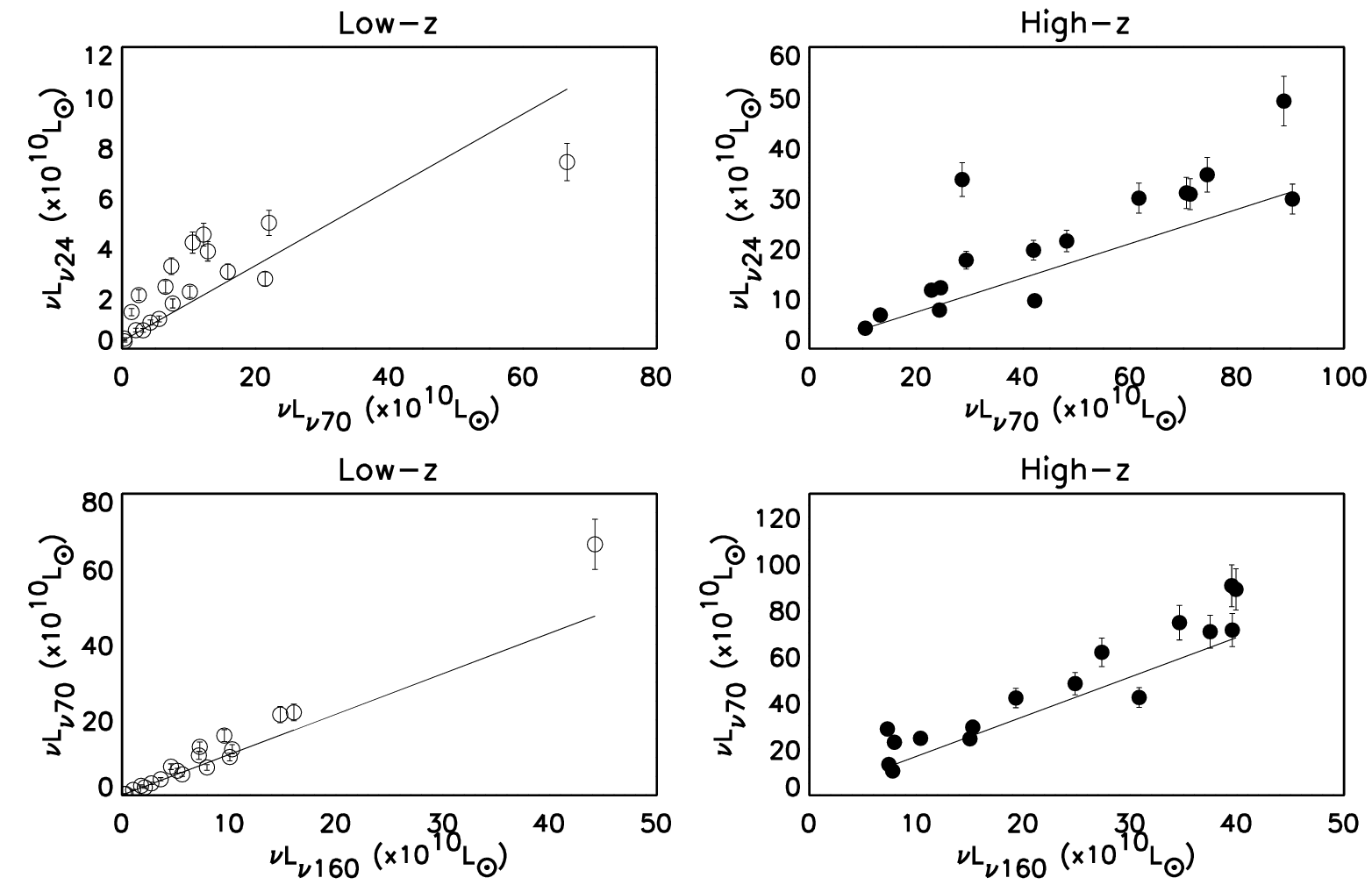}
\caption{Linear Mid-to-far infrared L$_{24}$ vs L$_{70}$ and L$_{70}$
vs L$_{160}$ luminosity correlations for the objects with
spectroscopic redshifts. The two redshift bins are $z \le 0.5$ (open
symbols, left) and $z>0.5$ (filled symbols, right). $\nu L_{\nu
\lambda}$ represents the monocromatic luminosity of the objects at
wavelength $\lambda$. The least square fit is shown for all four
cases.}
\label{mirfircorln}
\end{figure}


\begin{thebibliography}{}

\bibitem[Brand et al. (2006)]{brand06} Brand, K., et al., 2006, ApJ, 644, 143

\bibitem[]{}Bressan, A., Silva, L. $\&$ Granato, G., L., 2002, A$\&$A, 392, 377-391 

\bibitem[]{}Cram, L., et al., 1998, ApJ, 507, 155-160

\bibitem[]{}Coil, A., L., 2004, ApJ, 617, 765  

\bibitem[Dale \& Helou (2002)]{dale02} Dale, D.A., \& Helou, G., 2002, \apj, 576, 159

\bibitem[Dale et al. (2005)]{dale05} Dale, D.A., Bendo, G.J., Engelbracht, 
C.W., Gordon, K.D., et al., 2005, \apj, 633, 857

\bibitem[Fadda et al. (2002)]{dale05} Fadda, D., et al., 2002, A$\&$A, 383, 838 

\bibitem[Fazio et al.(2004)]{fazio04} Fazio, G.G., Ashby, M.L.N. et al., 
2004, \apjs, 154

\bibitem[Franceschini et al. (2003)]{fran03} Franceschini, A., Berta, S., Rigopoulou, D., Aussel, H., et al., 2003, A$\&$A, 403, 501

\bibitem[Frayer et al. (2006)]{frayer06} Frayer, D., T., et al., 2006, AJ, 131, 250

\bibitem[Genzel et al. (1998)]{genzel98}Genzel, R., Lutz, D., Sturm, E., Egami, E., 1998, \apj, 498, 579

\bibitem[Georgakakis et al. (2006)]{georg06} Georgakakis, A., et al., 2006, in press

\bibitem[Hogg et al. (2002)]{hogg02}Hogg, D., et al. 2002, astro-ph$/$0210394

\bibitem[Huang et al. (2004)]{huang04}Huang, J-S., et al., 2004, ApJS, 154, 44

\bibitem[]{} Kennicutt, R., C., 1998, ApJ, 272, 54-67

\bibitem[Neugebauer et al. (1984)]{neug84} Neugebauer, G., et al., 1984, ApJ, 278, 1

\bibitem[]{} Poggianti, B., M., Bressan, A.,  $\&$ Franceschini, A., 2001, ApJ, 550, 195-203

\bibitem[Puget et al.(1999)]{puget99} Puget, J.L., Lagache, G., 
Clements, D.L., Reach, W.T., et al., 1999, A$\&$A, 345, 29

\bibitem[Rieke et al. (2004)]{mips} Rieke, G.H., Young, E.T., et al., 2004,
\apjs, 154, 45 

\bibitem[Rigopoulou et al. (1999)]{rigo99} Rigopoulou, D., Spoon, H.,W.W., Genzel, R., et al., 1999 \aj, 118, 2625

\bibitem[]{} Rigopoulou, D., et al., 2000, ApJ, 537, L85-L89 

\bibitem [Rowan-Robinson $\&$ Crawford 1989]{}Rowan-Robinson, M., $\&$ Crawford, J., 1989, MNRAS, 238, 523-558

\bibitem [Sanders $\&$ Mirabel 1996]{}Sanders D. E. $\&$ Mirabel I. F., 1996, ARA$\&$A, 34, 749

\bibitem[Soifer et al. (1984)]{soifer} Soifer, B.T., Rowan-Robinson, M., 
Houck, J.R., de Jong, T., et al., \apj, 1984, 278, 71

\bibitem[Soifer et al. (1987)]{soifer} Soifer, B.T., et al., 1987, ARA$\&$A, 25, 187

\bibitem[Werner et al.(2004)]{wern04} Werner, M.W., 
Roellig, T.L., et al. 2004, \apjs, 254, 33


\end{thebibliography}
\end{document}